\documentclass[12pt,dvips]{article}
\usepackage[dvips]{graphicx}
\usepackage{graphicx}
\usepackage{latexsym}
\parskip=\itemsep               
\graphicspath{{plots/}}
\DeclareGraphicsExtensions{.eps.gz,.eps,.ps,.ps.gz}
\newcommand{\lwig}{\mbox{\,\raisebox{.3ex}
    {$<$}$\!\!\!\!\!$\raisebox{-.9ex}{$\sim$}\,}}
\newcommand{\gwig}{\mbox{\,\raisebox{.3ex}
    {$>$}$\!\!\!\!\!$\raisebox{-.9ex}{$\sim$}}\,}

\def\Journal#1#2#3#4{{#1} {#2} (#4) #3}

\def\NPB{{ Nucl. Phys.} B}
\def\PLB{{ Phys. Lett.}  B}

\def\PRD{{ Phys. Rev.} D}

\def\EPJC{{ Eur. Phys. J.} C} 
\def\RMP{ Rev. Mod. Phys.}
\def\CPC{ Comput. Phys. Commun.}
\def\NPS{ Nucl. Phys. Proc. Suppl.}

\def\be{\begin{equation}}
\def\ee{\end{equation}}
\def\bea{\begin{eqnarray}}
\def\eea{\end{eqnarray}}

\date{\empty}

\title{{\normalsize\rightline{DESY 00-172}\rightline{hep-ph/0012241}}
\vskip 1cm 
      \bf Zooming-in on Instantons at HERA  
       \vspace{21mm}} 
\author{A. Ringwald and F. Schrempp\\[4mm] 
Deutsches Elektronen-Synchrotron DESY, Hamburg, Germany}

\begin{document}
\begin{titlepage} 
  \maketitle
\begin{abstract}
In view of the intriguing, preliminary 
search results for instanton-induced events at HERA from the H1
collaboration, some important remaining theoretical issues are discussed.  
Notably, the question is addressed, to which extent the H1 analysis may be
directly compared to our original predictions from
instanton-perturbation theory, since certain
fiducial cuts are lacking in the H1 data. Various  
theoretical uncertainties are evaluated and their
impact on the observed excess is discussed.
An improved understanding of the experimental findings along with
an encouraging over-all agreement with our original predictions seems
to emerge.  
\end{abstract}


\thispagestyle{empty}
\end{titlepage}
\newpage \setcounter{page}{2}

{\it 1.} Instantons represent a basic non-perturbative aspect of QCD,
theoretically discovered and first studied by Belavin {\it et al.}~\cite{bpst}
and `t Hooft~\cite{th}, about 25 years ago.  
As topologically non-trivial fluctuations of the gauge fields with a
typical size of $0.3\,\div\,0.5$
fm~\cite{ukqcd,hasenfratz,stamatescu}, instantons play an important
r{\^o}le in the 
transition region between a partonic and a hadronic description of strong
interactions~\cite{ssh}. Yet, despite substantial theoretical evidence for the
importance of instantons in chiral symmetry breaking and hadron
spectroscopy,  their direct experimental verification is lacking until now.   
 
It turns out, however, that a characteristic {\it short distance}
manifestation of instantons can be exploited~\cite{rs1} for an experimental
search: Instantons induce certain (hard) processes that are forbidden in usual
perturbative QCD. These involve all (light) quark flavours
democratically  along with a violation of chirality, in accord
with the general chiral anomaly relation~\cite{th}.  

Deep-inelastic scattering (DIS) at HERA offers a unique
opportunity~\cite{rs1} to 
discover these hard processes induced by  QCD-instantons. It is of particular
importance that a theoretical prediction of both  the
corresponding rate~\cite{mrs,rs2,rs-lat} and
the characteristic event signature~\cite{rs1,cgrs,qcdins} is
possible in this hard scattering regime\footnote{For an exploratory
calculation of the instanton contribution to the gluon-structure
function, see Ref.~\cite{bb}.}. The instanton-induced cross section
turns out to be in a measurable  
range~\cite{rs2,rs-lat,rs-rev}, $\sigma^{(I)}_{\rm
HERA}\approx\mathcal{O}(30 \div 100)$ pb, depending on cuts. Crucial
information on the region of validity 
for this important result, based on instanton-perturbation theory, comes 
from a recent high-quality lattice simulation~\cite{rs-lat}. 
The main event signatures comprise a
``fireball''-like final state with a very high number 
of hadrons, including K mesons and $\Lambda$ hyperons, as well as a high
total transverse energy\footnote{A more extensive introduction to
instanton-induced events in DIS may e.g. be found in Ref.~\cite{rs-rev}.}.
With the help of the Monte Carlo generator QCDINS for
QCD-instanton-induced events in 
deep-inelastic scattering~\cite{qcdins}, the H1 and ZEUS experiments at HERA 
are actively searching for signatures of instantons in the hadronic
final state. The challenging experimental task is to distinguish the
instanton-induced signal from the normal DIS final state near the edge of the
available phase space. This effort seems however well worthwhile, 
since an experimental verification of such a novel,
non-perturbative manifestation of QCD would be of basic
significance.

At the recent DIS2000 and ICHEP2000 conferences, the H1 collaboration
has reported preliminary results of a first dedicated search for
instanton-induced events at HERA~\cite{mikocki,h1_ichep}. 
In a phase space region, where a reduction of the normal DIS (nDIS)
background to the percent level is achieved according to standard Monte Carlo
models, a (statistically) significant excess of events was found in
the H1 data. 
While its size is at a level still comparable to the differences among 
standard DIS event generators, it is - for the discriminating observables -
qualitatively similar to the expected instanton signal. 
The results presented are quite intriguing  and encouraging, although
far from being conclusive. Yet, they strongly enhance the  motivation for
looking more closely at some remaining theoretical issues related to
our original predictions. 

\begin{figure}[ht]
\begin{center}
\begin{tabular}{ll}
\mbox{\includegraphics*[width=7.5cm]{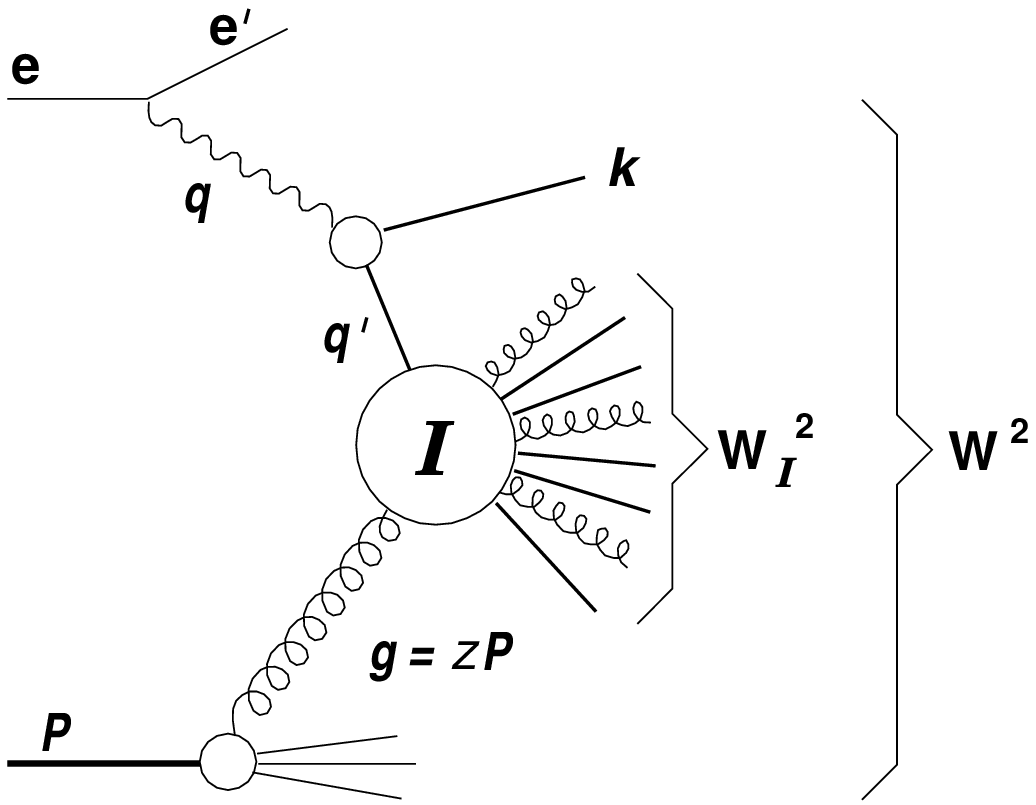}}
&
{\small
\renewcommand{\arraystretch}{1.0}
\begin{tabular}{l}
 \vspace{-5.5cm} \\
 DIS variables: \\[1ex]
   $S=(e+P)^2$\\
   $Q^2 = - q^2 = -(e-e^\prime\,)^2$\\ 
   $ x_{\rm Bj} = Q^2 / \; (2 P \cdot q) $ \\
   $ W^2 =(q+P)^2 = Q^2 (1/x_{\rm Bj}-1)$ \\[1ex]
   $ x = Q^2/\; (2 g\cdot q) = x_{\rm Bj}/\;z$\\[2ex]
 Variables of instanton-subprocess: \\[1ex]
 $Q^{\prime\,2}=- q^{\prime\,2} = - (q-k)^2 $ \\
 $x^\prime = Q^{\prime\,2}  / \;(2 \; g \cdot q^\prime ) $ \\
 $W_I^2= (q^\prime+g)^2 = Q^{\prime\,2} ( 1/x^\prime -1)$\\
\end{tabular}}
\end{tabular}
\end{center}
\vspace{0.3cm}
   \caption[]
     {Structure and kinematic variables of the dominant
     instanton-induced process in deep-in\-elastic scattering.} 
   \label{kin-var}
\end{figure}

{\it 2.}  Let us start off by briefly summarizing the strategy and
essential results of the H1 search for instanton-induced
events~\cite{mikocki,h1_ichep}. Structure and kinematical variables of
the dominant instanton-induced process in deep-in\-elastic scattering
are displayed in Fig.~\ref{kin-var}. Of particular importance will be
the Bjorken-variables ($Q^{\prime\,2},x^\prime $) of the
so-called instanton subprocess $q^\prime+g\Rightarrow I\,\Rightarrow  X$. 

The H1 analysis is based on the strategy~\cite{cgrs}
of isolating an ``instanton-enriched'' data sample by means of
suitable cuts to a set of three sensitive, discriminating
observables, defined in 
Table~\ref{obs} (left). Three different cut-scenarios A), B) and C)
with {\it increasing} instanton-separation power 
$\epsilon_{\rm I}/\epsilon_{\rm nDIS}$ were defined
according to
\begin{itemize}
\itemsep 0.2ex
\item[A)] the highest instanton efficiency ($\epsilon_{\rm I}$);
\item[B)] high $\epsilon_{\rm I}$ at reasonable normal DIS background reduction
(normal DIS efficiency $\epsilon_{\rm nDIS}$); 
\item[C)] highest instanton-separation power at $\epsilon_{\rm I}\approx 10\%$.
\end{itemize}

\vspace{2ex}
\begin{table} [b]
\begin{center}
\begin{tabular}{|ll|ll|}\hline 
\multicolumn{4}{|c|}{{\bf Observables}}\\\hline\hline
$n_{\rm ch,\,b}$ = &number of charged particles&$Et_{\rm
b}$ = &total transverse energy\\
&in ``instanton band''&&in ``instanton band''\\\hline 
$\rm Sph$ = &sphericity in the rest system of& $1-E_{\rm out\,,b}/E_{\rm
in\,,b}$ = &$Et$-weighted azimuthal \\
&particles not from current jet&&event isotropy \\\hline
$Q^{\prime\,2}_{\rm rec}$ = &reconstructed virtuality&$Et_{\rm jet}$ =
&transverse energy of\\
&of quark entering $I$-subprocess &&current jet\\\hline  
\end{tabular}
\caption[dum]{Set of discriminating instanton-sensitive observables as
used by the H1
collaboration~\cite{mikocki,h1_ichep}. Instanton-enhancement cuts are
applied to the three observables on the left. 
All observables
refer to the hadronic CMS ($\vec{q}+\vec{P} = \vec{0}$), except for
the sphericity that is calculated in the rest system of the particles
not belonging to the current jet. The latter is identified with the
jet of highest $Et$. The ``instanton band'' is a prominent theoretical
prediction~\cite{rs1,rs-rev} and is experimentally placed at
$\overline{\eta}\pm 1.1$ where $\overline{\eta}$ is the $Et$-weighted
mean pseudorapidity of all particles not belonging to the current
jet.\label{obs}} 
\end{center}
\end{table}
In all three cut-scenarios, significantly  more events are observed
than expected by standard DIS Monte Carlo models. It is noteworthy
that with increasing instanton-separation power an increasingly large excess is
seen in the data. The H1 collaboration has compared the observed
excess in each of the six observables (Table~\ref{obs}) to our
predictions from the QCDINS 2.0 Monte Carlo 
generator~\cite{qcdins}. It is found that the shape and the size of
the excess in 
the three discriminating observables $n_{\rm ch,\,b},\ {\rm Sph}$ and
$Q^{\prime\,2}_{\rm rec}$ with instanton-enhancement cuts, as well as
in the azimuthal event isotropy $1-E_{\rm out\,,b}/E_{\rm
in\,,b}$, are in qualitative agreement with the expected instanton signal.    
However, the comparison with the two remaining observables, $Et_{\rm
b}$ and $Et_{\rm jet}$, is more involved. While there appears to be
also a marked excess in the respective data, both $\langle Et_{\rm
b}\rangle$ and $\langle Et_{\rm jet}\rangle$ appear to be shifted
towards smaller values than predicted by QCDINS, and the widths of the
experimental distributions are also considerably narrower.

\begin{figure}
\vspace{2ex}
\begin{center}
\begin{tabular}{|lc|c|c|}\hline
        && Instanton
        Events&\rule[-1.5ex]{0ex}{5ex}$
        \Lambda^{(n_f=3)}_{\overline{\rm MS}}$ \\   
\multicolumn{2}{|c|}{\raisebox{1.0ex}[-1.0ex]{\bf Cuts} }& ($ \int
\mathcal{L} dt =15.8 {\rm\
pb}^{-1}$)&\rule[-1.5ex]{0ex}{3.5ex}[MeV]\\\hline\hline    
\rule[-2ex]{0ex}{5ex}fiducial (theory):& $x^\prime\ge 0.35,\
Q^\prime/\Lambda^{(n_f)}_{\overline{\rm MS}}{\ge 30.8}$&&\\\cline{1-2}
 &\rule[0ex]{0ex}{3ex} $x_{\rm Bj}\ge 10^{-3},\ 0.1\le y_{\rm Bj}\le 0.6$,&&\\
\raisebox{1.0ex}[-1.0ex]{kinematical:}&\rule[-2ex]{0ex}{2.5ex}$\theta_{el}
> 156^{\,\circ},\ E^\prime_{el} \ge 10$ GeV&\raisebox{3.0ex}[-3.0ex]
{$633{{-154}\atop{+194}}$}&\raisebox{1.0ex}[-1.0ex]
{$346{{+31}\atop{-29}}$}\\\cline{1-3}   
instanton enhancement:&\rule[0ex]{0ex}{3ex} $n_{\rm ch,\,b} \ge 8,\
{\rm Sph} > 0.5$,& &\\ 
(cut-scenario
{\bf C}~\cite{mikocki,h1_ichep})&\rule[-2ex]{0ex}{2.5ex}$105<Q^{\prime\,2}_{\rm
rec} < 200 {\rm\ GeV}^2$ &\raisebox{1.0ex}[-1.0ex]{$163{{-23}\atop{
-28}}$}&\\\hline
\end{tabular}

\vspace{1.3ex} 
\parbox{16cm}{\includegraphics*[angle=-90,width=15.cm]{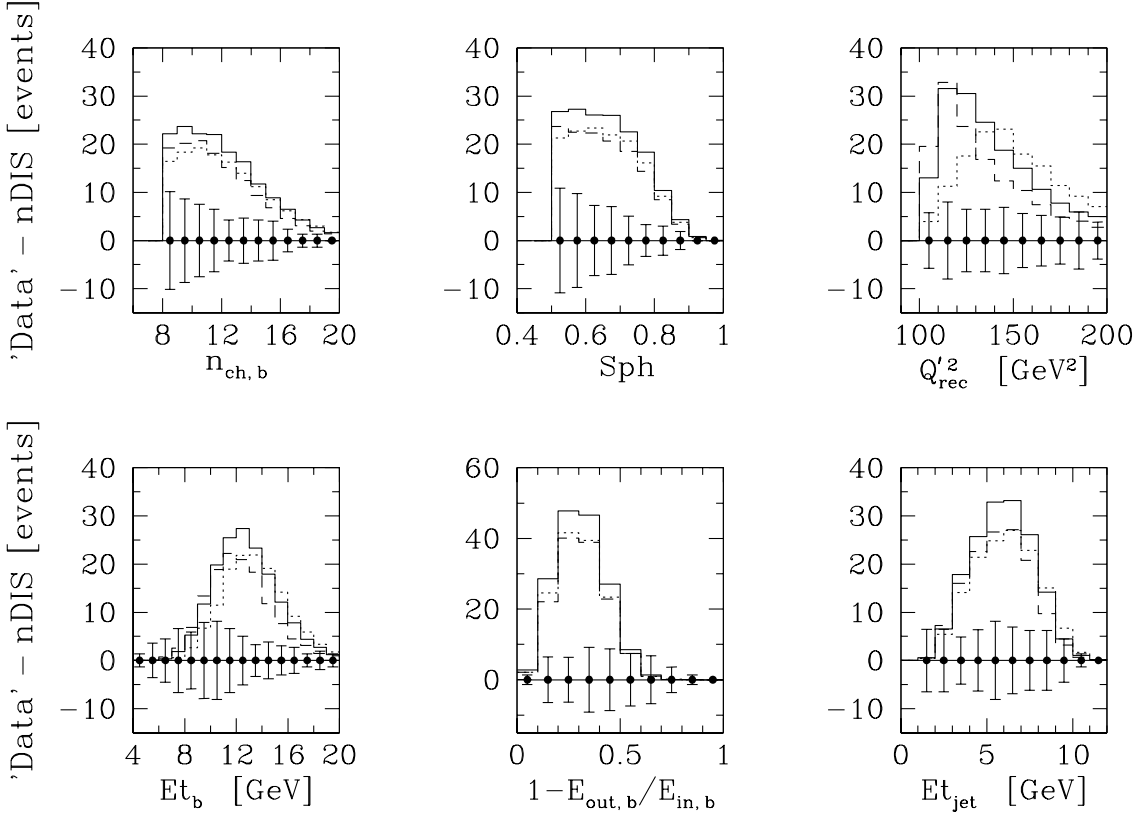}}
\caption[dum]{Predicted excess due to instanton-induced events (QCDINS
2.0~\cite{qcdins}), relative to the normal DIS expectations (nDIS). 
Except for missing detector corrections, all cuts, luminosity
(c.\,f. Table above), observables (c.\,f. Table~\ref{obs}) 
and the parton densities~\cite{cteq} are as for the preliminary H1
data~\cite{mikocki,h1_ichep}. While refraining to show the  
real data before their final publication, we have instead
displayed the statistical errors as calculated by means of the
HERWIG 5.9~\cite{herwig} event generator for vanishing excess. 
Apart from modelling the achievable significance of the instanton
signal via a good approximation of the actual (statistical) 
errors, the figure displays the dependence of the various observables
on the uncertainty of $\Lambda^{(3)}_{\overline{\rm MS}}$. The solid,
dotted and dashed lines refer to the 1998 world average value
$\Lambda^{(5)}_{\overline{\rm MS}} = 219$ MeV~\cite{pdg98}, its upper
and lower $1\sigma$ limits, respectively.  
\label{h1-obs}} 
\end{center}
\end{figure}

Figure~\ref{h1-obs} may serve as an illustration of 
the achievable instanton signature in the set of six
instanton-sensitive observables (c.\,f. Table~\ref{obs}) with cuts, integrated
luminosity ($\int \mathcal{L} dt =15.8 {\rm\ pb}^{-1}$) and parton
densities~\cite{cteq}  as for the preliminary H1 data~\cite{mikocki,h1_ichep}. 
While refraining to show the real data before their final publication,
we have instead displayed the statistical errors as calculated by means of the
HERWIG 5.9~\cite{herwig} event generator for vanishing excess. On the
one hand, the corresponding number of $361\pm 26$ nDIS events from HERWIG 
after instanton-enhancement cuts (cut-scenario {\bf C}~\cite{mikocki,h1_ichep})
matches quite well the values quoted by the H1 collaboration. On the
other hand, the detector corrections (not accounted for here) seem to
significantly reduce the number of instanton events by almost a factor
of two.

A central issue to be discussed in this letter is
the question to which extent the H1 analysis may be directly compared to our
theoretical predictions, since certain fiducial cuts are lacking in
the H1 data: Because of difficulties with the $x^\prime$-reconstruction, 
there is presently no cut on
$x^\prime$, while the data are compared to QCDINS with the default
$x^\prime$-cut implemented. Moreover, both the data and the QCDINS
results used are lacking the theoretically required cut in $Q^2 \gwig
\mathcal{O}(100)$ GeV$^2$. We also use this opportunity to
reconsider in some detail the theoretical uncertainties due to the
extraction of the fiducial cuts in $(Q^{\prime\,2},x^\prime)$ from
lattice data and due to the known uncertainties $\delta
\Lambda_{\overline{\rm MS}}$ of the QCD scale.

{\it 3.} Proceeding in reverse order for reasons of presentation, let
us turn first to the uncertainties arising from $\delta 
\Lambda_{\overline{\rm MS}}$  via the known strong
dependence~\cite{rs2} of the instanton subprocess cross section (c.\,f.
Fig.~\ref{kin-var}) on the QCD scale parameter,
\begin{equation}
\sigma^{(I)}_{q^\prime g}\, (Q^{\prime\,2},x^\prime ) \propto \left(
\frac{\Lambda^2_{\overline{\rm MS}}}{Q^{\prime\,2}}\right)^{\beta_0
F(x^\prime)};\hspace{2ex} \beta_0 = 11 -\frac{2}{3}n_f;\hspace{2ex}
1/2 \leq F(x^\prime)\leq 1. 
\label{lamdep}
\end{equation}   
The dependence of the various observables on the uncertainty of
$\Lambda^{(n_f=3)}_{\overline{\rm MS}}$ is displayed in Fig.~\ref{h1-obs}. The
solid line corresponds to the default prediction of our
QCDINS event generator~\cite{qcdins}, with
$\Lambda^{(3)}_{\overline{\rm MS}} = 346{{+31}\atop{-29}}$ MeV,
being obtained  by ``flavour reduction'' with 3-loop accuracy from the
1998 world-average value~\cite{pdg98} $\Lambda^{(5)}_{\overline{\rm MS}} = 219
{{+25}\atop{-23}}$ MeV or $\alpha_s(M_Z)= 0.119\pm 0.002$. Note that
the errors quoted for the total expected instanton-event numbers in
Fig.~\ref{h1-obs} only refer to these $\Lambda^{(5)}_{\overline{\rm
MS}}$ uncertainties.  The optimization of the
instanton-enhancement cuts (cut-scenario {\bf C} of
Ref.~\cite{mikocki,h1_ichep}) reflects in the number of instanton events
being maximal for the default value of $\Lambda^{(5)}_{\overline{\rm MS}}$.

Apparently, the actual dependence of the various observables on 
$\delta\Lambda^{(3)}_{\overline{\rm MS}}$ is much less dramatic than naively
expected from Eq.~(\ref{lamdep}). The reason is, of course, associated
with the fact that the fiducial $Q^{\prime}$-cut in Fig.~\ref{h1-obs} is
proportional to $\Lambda^{(3)}_{\overline{\rm MS}}$. This is a natural
consequence from exploiting lattice results to which we shall turn next. 

{\it 4.} Let us reconsider~\cite{rs2,rs-lat} in more detail the extraction
of the fiducial cuts in ($Q^{\prime\,2},x^\prime$) from a recent
high-quality lattice simulation of quenched QCD~\cite{ukqcd}, 
with special emphasis on the major associated uncertainties\footnote{    
A proper account of the considerable systematic uncertainties
associated with different cooling/smoothing methods in the lattice 
simulations~\cite{ukqcd,stamatescu} and \cite{hasenfratz} is very hard and
clearly beyond the scope of this paper.}.
 
\begin{figure} [tb] 
\vspace{-3ex}
\begin{center}
\parbox{6cm}{\includegraphics*[width=6cm]{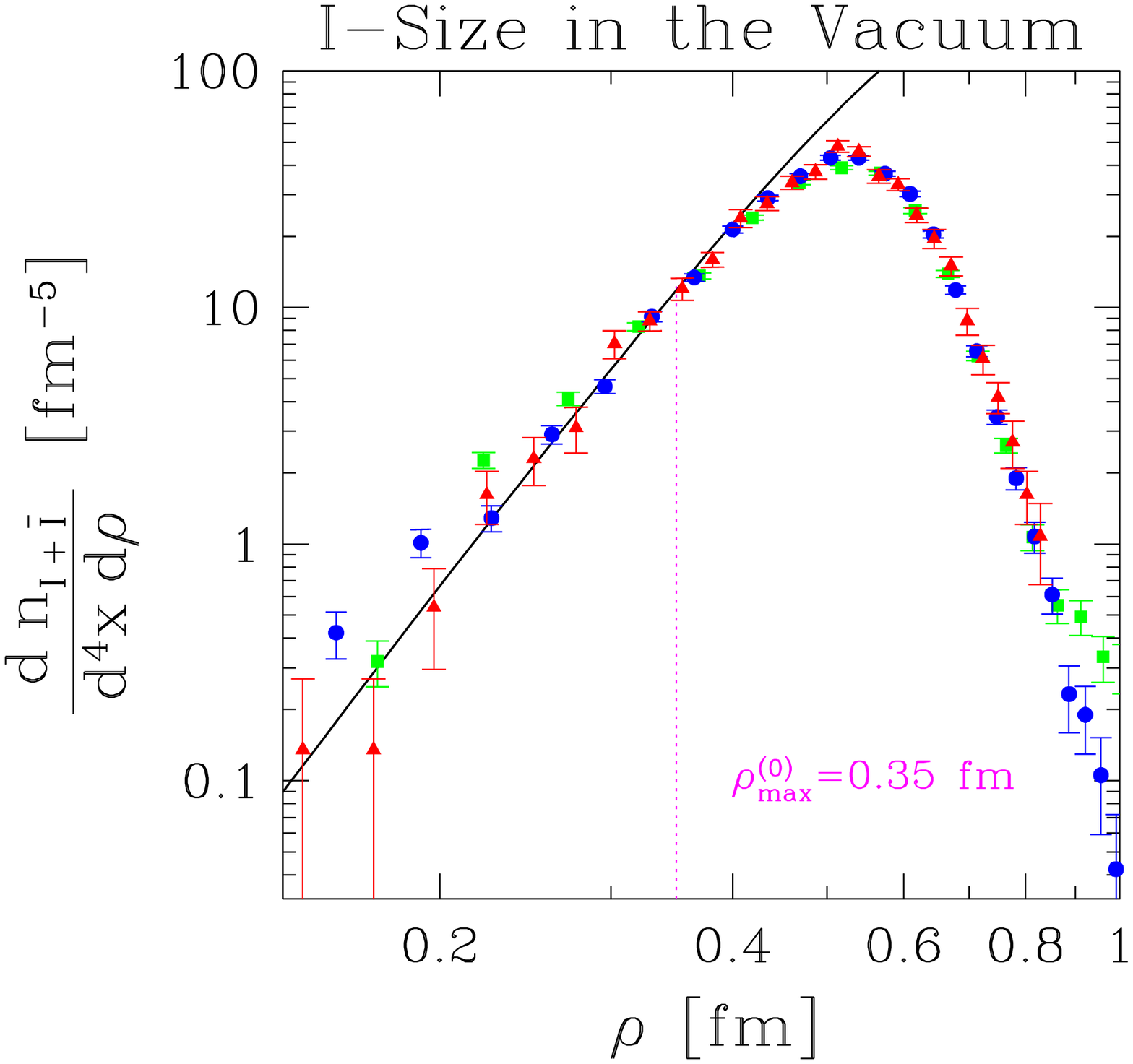}}
\hspace{10ex}\parbox{6cm}{\includegraphics*[width=6cm]{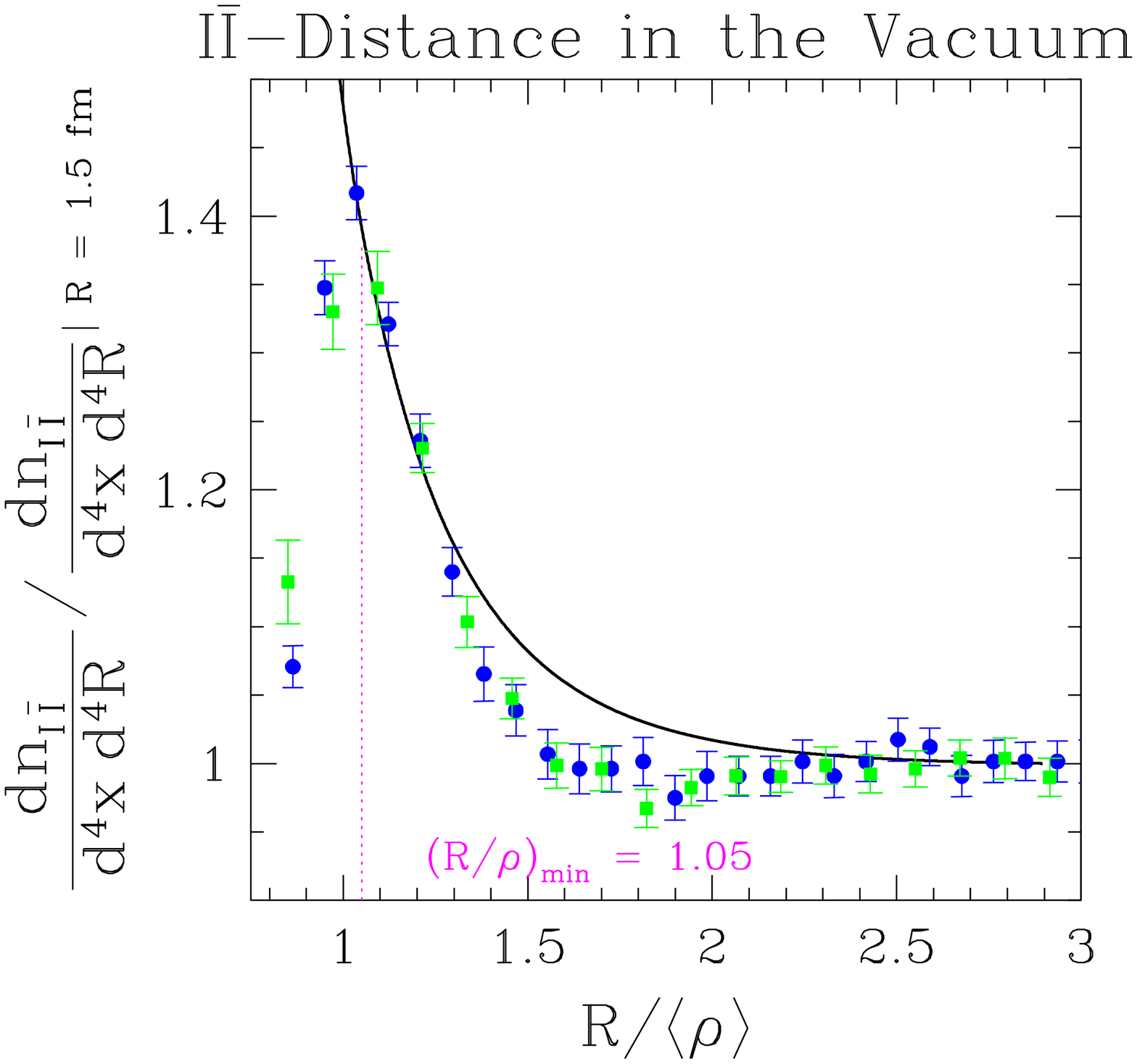}}\\[2ex]
\hspace{-1.5ex}\parbox{6cm}{\includegraphics*[width=6cm]{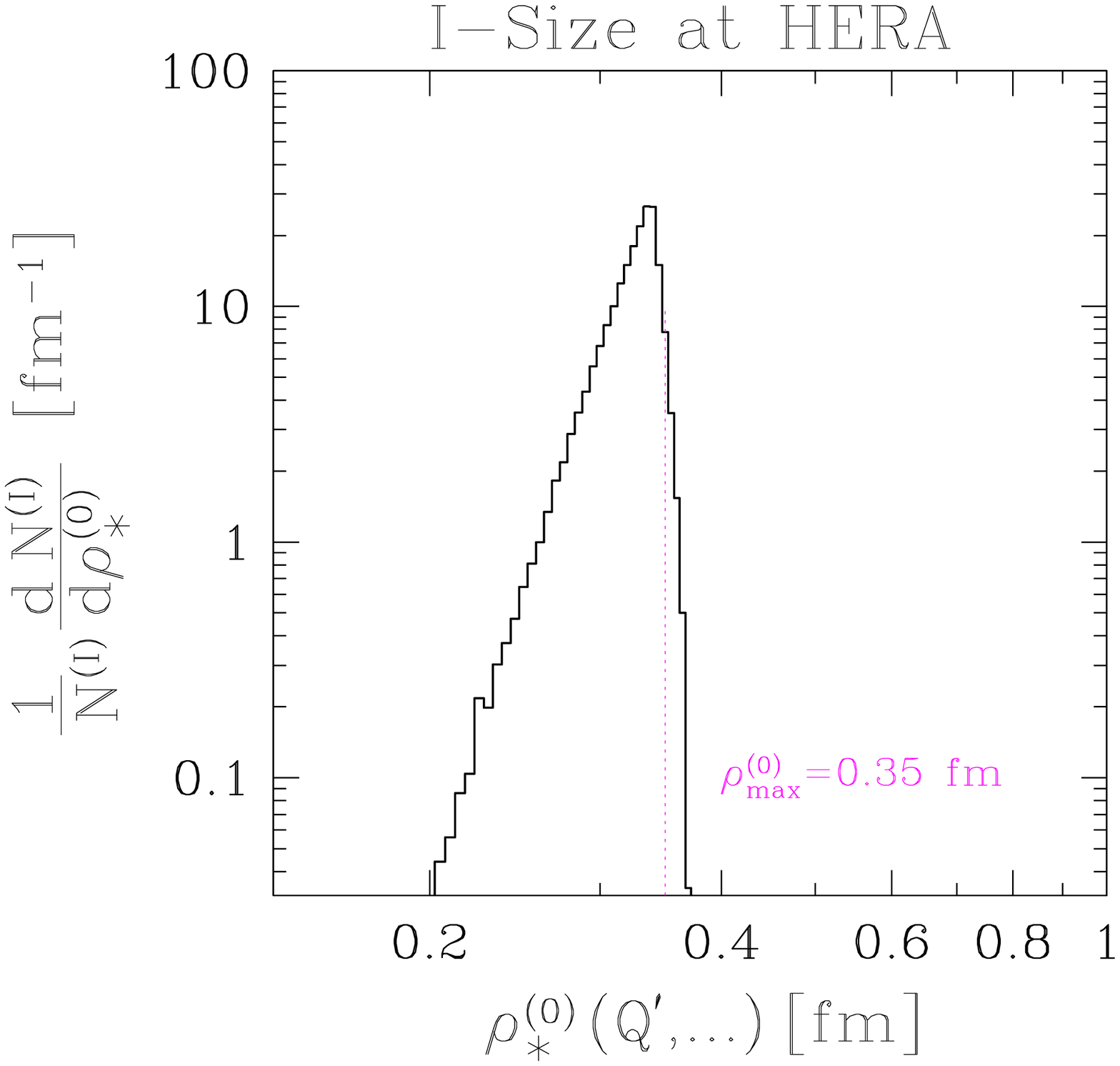}}
\hspace{8.95ex}\parbox{6cm}{\includegraphics*[width=6cm]{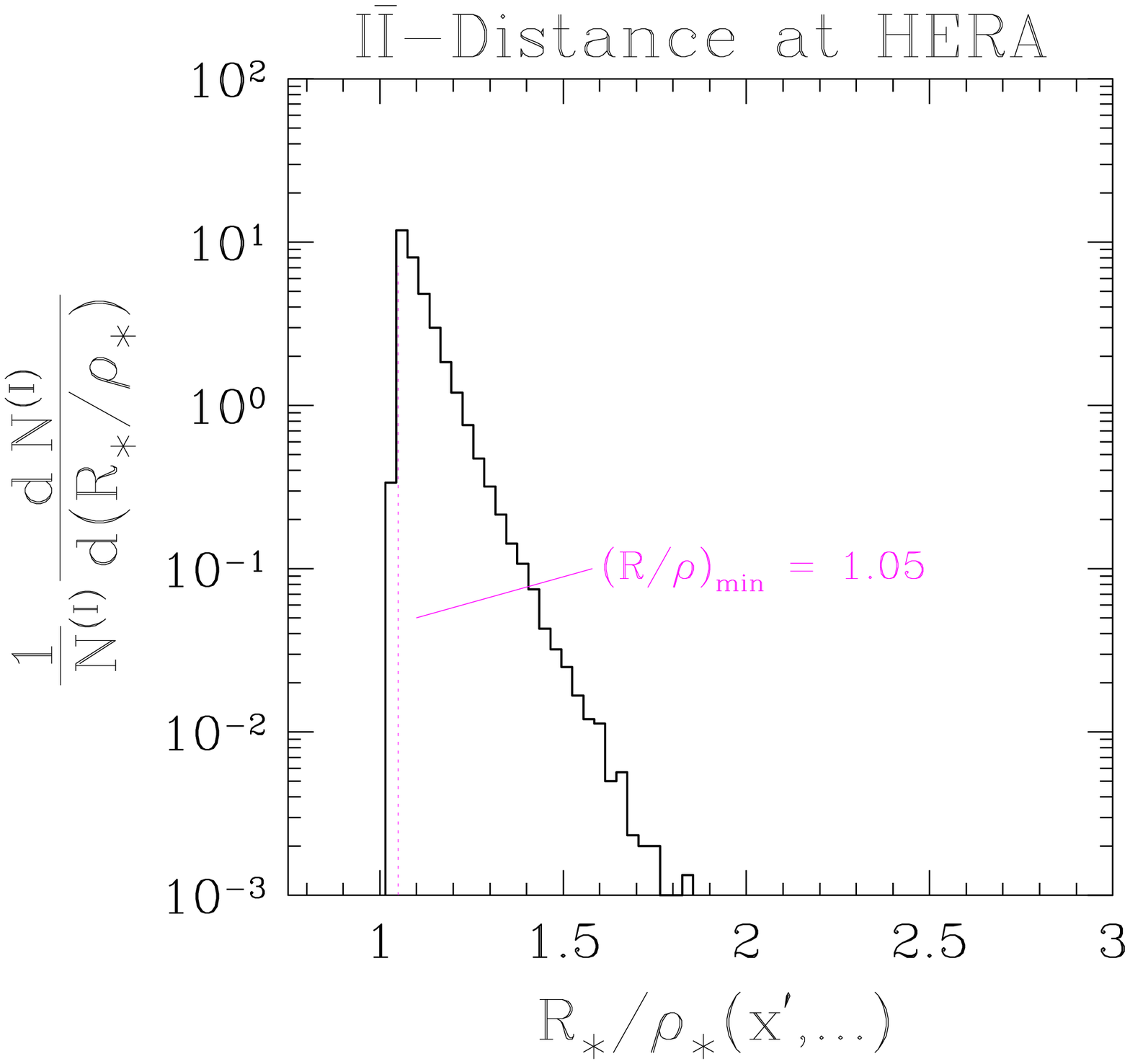}}
\caption[dum]{Top: Illustration of the agreement
of recent high-quality lattice data~\cite{ukqcd,rs-lat} for the
instanton-size distribution and the $I\overline{I}$-distance
distribution with the predictions from  
instanton-perturbation theory~\cite{rs-lat} for $\rho\lwig
0.35$ fm and $R/\rho\gwig 1.05$, respectively. 
Bottom: Display of the instanton-induced event distributions for HERA
from QCDINS, as function of the saddle-point values
($\rho_\ast(Q^\prime,\ldots),\,R_\ast/\rho_\ast(x^\prime,\ldots)$),
with the default $(Q^{\prime\,2},\,x^\prime)$-cuts implemented as in
Fig.~\ref{h1-obs}. In these variables, the above cut-off values
$(\rho^{(0)}_{\rm max},\ (R/\rho)_{\rm min})$ extracted from the
lattice are apparent.\label{lattice}}  
\end{center}
\end{figure}

The validity of instanton-pertur\-bation theory, on which our
predictions for HERA are based, requires instantons of small enough
size $\rho\le\rho_{\rm max}$ along with a sufficiently 
large separation $R/\rho\ge (R/\rho)_{\rm min}$ between them. Crucial
information on $(\rho_{\rm max},\ (R/\rho)_{\rm min})$ was
obtained~\cite{rs-lat} by  confronting the predictions of
instanton-perturbation theory with the UKQCD lattice
``data''~\cite{ukqcd} for the 
$I$-size distribution $\frac{dn_I}{d^4x\,d\rho}$ and the
$I\overline{I}$-distance distribution $\frac{d n_{I\overline{I}}}{d^4
x\,d^4 R}$ (c.f. Fig.~\ref{lattice} (top)).  

Let us recall~\cite{rs2,rs-lat} that the instanton-induced cross section
$\sigma^{(I)}_{\rm HERA}$ depends strongly on these lattice observables
via Fourier-type transformations\footnote{In
Eq.~(\ref{cs}), $D$, $\Omega$ and $\{\ldots\}$ denote the known
theoretical expressions for the $I$-size distribution, the 
$I\overline{I}$-valley interaction and further smooth factors, respectively.},
\begin{equation}
\sigma_{q^\prime g}^{(I)}(Q^{\prime\,2},x^\prime) \sim 
      \int d^4 R \ {\rm e}^{{\rm i} (p+q^\prime)\cdot R}
	\underbrace{
      \int\limits_0^\infty d\rho 
      \int\limits_0^\infty d\overline{\rho}\,
      \underbrace{D(\rho)}_{\frac{dn_I}{d^4x\,d\rho}}
      D(\overline{\rho}) 
       \int dU
            {\rm e}^{{-\frac{4\pi}{\alpha_s}}
      \Omega\left(\frac{R^2}{\rho\overline{\rho}},
      \frac{\overline{\rho}}{\rho},U \right)}}_{\frac{d
      n_{I\overline{I}}}{d^4 x\,d^4 R}}
       \ {\rm e}^{-Q^\prime(\rho+\overline{\rho})}
      \ \{\ldots \}\, .
\label{cs}
\end{equation}
Due to this structure,
the limits ($\rho_{\rm
 max},\ (R/\rho)_{\rm min}$) for the validity of
 instanton-perturbation theory from lattice data may, in fact, be translated in
 a  one-to-one  manner, via a saddle-point relation,   
to {\it minimally required cuts} on the conjugate Bjorken variables
($Q^{\prime\,2},x^\prime$) of the instanton-subprocess as follows,
\begin{equation}
\left.
\begin{array}{r} 
 \frac{Q^\prime_{\rm min}}{\Lambda^{(n_f)}_{\rm \overline{MS}}}\ \\[4ex]
x^\prime_{\rm min}\
\end{array}\right\}
\hspace{1ex}\stackrel{\rm saddle-point}{\Leftrightarrow}\hspace{1ex}
\left\{ \begin{array}{lcr}
\ \frac{\beta^{(n_f)}_0}{\Lambda^{(n_f)}_{\rm
 \overline{MS}}}\,\frac{1}{\rho^{(n_f)}_{\ast\,{\rm max}}}
&\approx\ &  \frac{\beta^{(0)}_0}{\Lambda^{(0)}_{\rm 
 \overline{MS}}}\,\frac{1}{\rho^{(0)}_{{\rm max}}};\nonumber\\[4ex]
\ \left(\frac{R_\ast}{\rho_\ast}\right)^{(n_f)}_{\rm min} 
&\approx\ &\left(\frac{R}{\rho}\right)^{(0)}_{\rm min}.
\label{saddle}
\end{array}\right.
\end{equation}
According to the known $n_f$-dependence of the solutions
($\rho_\ast,R_\ast$) of the saddle-point
equations~\cite{rs2}, one finds that the combinations ($\rho_\ast\,\Lambda_{
\overline{\rm MS}}/\beta_0, R_\ast/\rho_\ast$) are approximately 
$n_f$-independent functions of ($Q^{\prime}/\Lambda_{\overline{\rm
MS}},x^\prime$). In addition, Eq.~(\ref{saddle}) incorporates the working
hypothesis that 
the $n_f$-independence of this combination extends also to the
corresponding limits of validity of instanton-perturbation theory.
With high-quality lattice data being only available for $n_f=0$ at
present, this appears to be a near-at-hand prescription to account for effects
of light flavours in the required fiducial cuts.
In practice, one will identify the actual
($Q^{\prime},x^\prime$) cuts with the ``fiducial'' ones, i.e.
($Q^{\prime}_{\rm min},x^\prime_{\rm min}$), in order to profit from
a high event rate without loosing theoretical control.  

Before turning to a discussion of uncertainties associated with this
procedure, let us briefly recapitulate the main result on the fiducial region
from Ref.~\cite{rs-lat} and its implementation in the
QCDINS~\cite{qcdins} event generator for HERA. 
First of all, Fig.~\ref{lattice} (top) illustrates the good agreement
of the data for the instanton-size distribution
and the $I\overline{I}$-distance distribution from
Refs.~\cite{ukqcd,rs-lat} with the predictions from 
instanton-perturbation theory from Refs.~\cite{rs-lat,qcdins}, for  
\begin{equation}
 \left.\begin{array}{lcccl}\rho& \lwig &  
         \rho^{(0)}_{\rm max}&\approx & 0.35 {\rm\ fm};\\[1ex]
 \frac{R}{\rho}& \gwig& \left(\frac{R}{\rho}\right)_{\rm
 min}&\approx & 1.05;  \\
 \end{array}\right\}\hspace{2ex}{\rm implying}\hspace{2ex}
 \left\{\begin{array}{lcl}Q^\prime/\Lambda^{(n_f)}_{\overline{\rm MS}} 
&\gwig &
 30.8;\\[1ex]
 x^\prime& \gwig &0.35.\\
 \end{array} \right .
\label{fiducial}
\end{equation}
Figure~\ref{lattice} (bottom) displays  the
instanton-induced event distribution for HERA from QCDINS, as function of the
saddle-point values
($\rho^{(0)}_\ast(Q^\prime,\ldots),R_\ast/\rho_\ast(x^\prime,\ldots)$),
with default fiducial cuts in
($Q^{\prime\,2},x^\prime$) as in Fig.~\ref{h1-obs}
(top). Apparently, the event distribution correctly 
reflects the above cut-off values $(\rho^{(0)}_{\rm max},\ (R/\rho)_{\rm min})$
extracted from the lattice. 

The next issue is to extract a more quantitative estimate for the
associated uncertainties. Starting with the $\rho$-dependence in the
$I$-size distribution, we  also take into account the error on the
QCD-scale parameter $\Lambda^{(0)}_{\overline{\rm
MS}}$ for $n_f=0$ from Ref.~\cite{alpha} as follows. We treat
$\Lambda^{(0)}_{\overline{\rm 
MS}}$ as a free parameter fluctuating around its
quoted mean value, $\langle\Lambda^{(0)}_{\overline{\rm
MS}}\rangle=238$ MeV, with error $ \delta \Lambda^{(0)}_{\overline{\rm
MS}}=19$ MeV, in the following $\chi^2$ function, 
\begin{equation}
\chi^2(\Lambda^{(0)}_{\overline{\rm MS}}\rho,s)
=\min_{\Lambda^{(0)}_{\overline{\rm MS}}}\,\left\{ 
\sum_{\rho_{i}\le 
\rho}\left(\frac{\frac{dn_{I+\overline{I}}(\rho_i)}{d^4x\,d\rho}-
2\,D_{I-pert.th.}(\rho_i,s,\Lambda^{(0)}_{\overline{\rm
MS}})}{\delta\,\frac{dn_{I+\overline{I}}(\rho_i)}{d^4x\,d\rho}}\right)^2
+\left(\frac{\Lambda^{(0)}_{\overline{\rm
MS}}-\langle\Lambda^{(0)}_{\overline{\rm
MS}}\rangle}{\delta\Lambda^{(0)}_{\overline{\rm
MS}}}\right)^2\right\}. 
\label{chi2min}
\end{equation}
\begin{figure}
\vspace{-1ex}
\begin{center}
\parbox{6cm}{\includegraphics*[width=6cm]{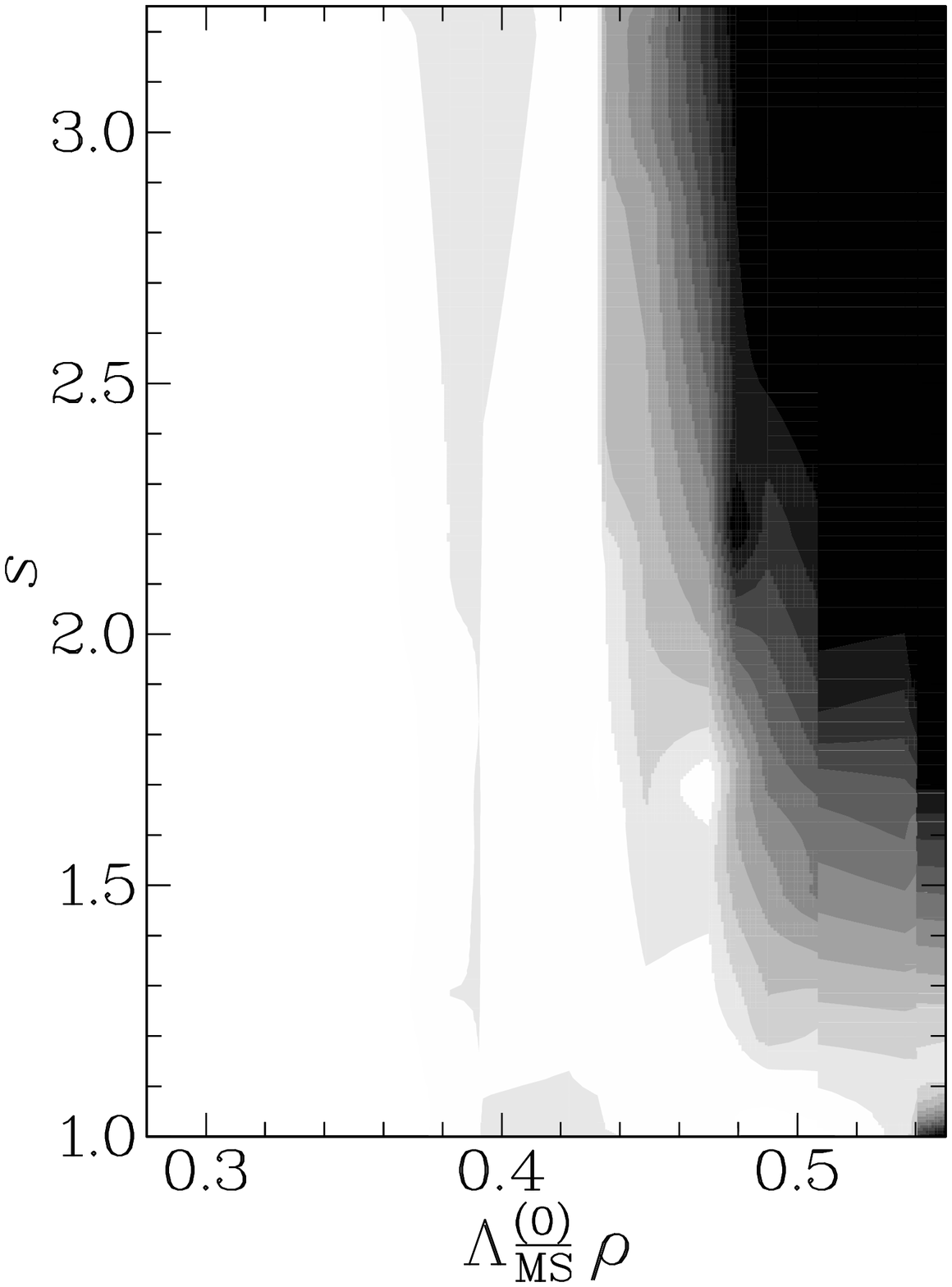}}\hspace{8ex}
\parbox{6.5cm}{\includegraphics*[width=6.5cm]{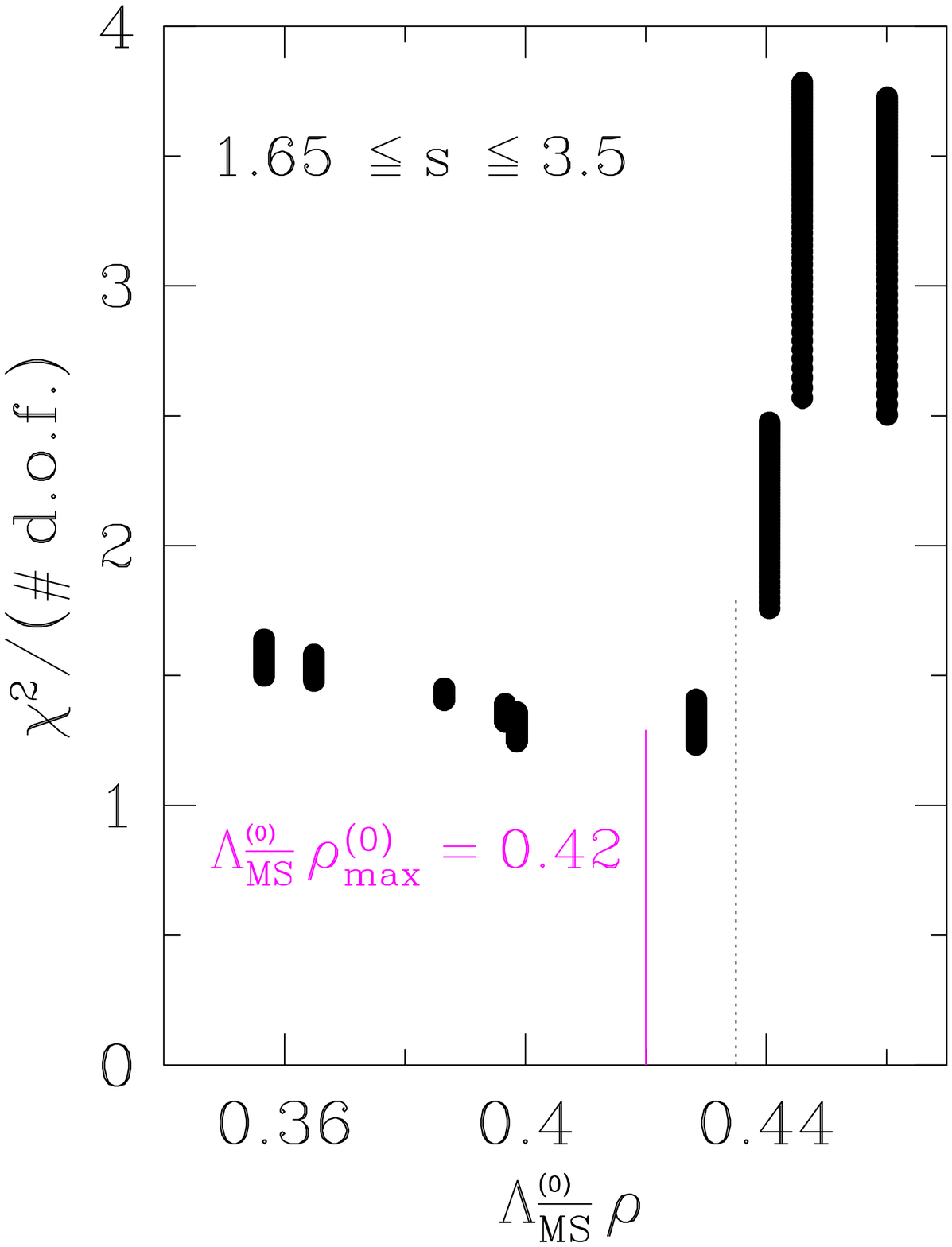}}
\caption[dum]{$\chi^2/(\#\,{\rm d.o.f.})$,
characterizing the deviation of instanton-perturbation theory from
the UKQCD lattice data~\cite{ukqcd,rs-lat} for the $I$-size
distribution according to 
Eq.~(\ref{chi2min}). Left: Contour plot for a range of 
maximal values of $\Lambda^{(0)}_{\overline{\rm MS}}\,\rho$ 
and of the renormalization-scale factor $s$.  Darker shading corresponds to
increasing $\chi^2/(\#\,{\rm d.o.f.})$.  Right: Projection on the
$\Lambda^{(0)}_{\overline{\rm MS}}\,\rho$-axis for a range of larger
$s$-values. The solid line corresponds to the default
$\Lambda^{(0)}_{\overline{\rm MS}}\,\rho$ cut-off in QCDINS 2.0 and
the dotted one to an estimate of the associated uncertainty.    
\label{chi2}} 
\end{center}
\end{figure}

The theoretical expression  for the
$I$-size distribution
$D_{I-pert.th.}(\rho_i,s,\Lambda^{(0)}_{\overline{\rm MS}})$ from
instanton-per\-tur\-ba\-tion theory is taken as in
Ref.~\cite{rs-lat}. It incorporates renormalization-group invariance
at the two-loop level, $d\, \ln(
D(\rho,s,\Lambda^{(0)}_{\overline{\rm MS}}))/\,d\,\ln(s)
=\mathcal{O}(\alpha_s^2)$, with the renormalization-scale dependence
parametrized conveniently as 
\begin{equation}
\mu = s/\rho.
\end{equation} 
The resulting $\chi^2/(\#\,{\rm d.o.f.})$, minimized with respect
to $\Lambda^{(0)}_{\overline{\rm MS}}$, is displayed in
Fig.~\ref{chi2} for a range of
maximal values of $\Lambda^{(0)}_{\overline{\rm MS}}\,\rho$ 
and of the renormalization-scale factor $s$.  
First of all, it is important to consider sufficiently large
$s>1$, where the incorporated two-loop renormalization-group invariance 
should lead to a largely $s$-independent limit $\rho^{(0)}_{\rm max}$.
Indeed, this is nicely reflected in the contour-plot of the resulting
$\chi^2/(\#\,{\rm d.o.f.})$ versus $s$ and 
$\Lambda^{(0)}_{\overline{\rm MS}}\,\rho$ in Fig.~\ref{chi2} (left). 
The projection on the  $\Lambda^{(0)}_{\overline{\rm MS}}\,\rho$-axis
for $1.65\le s\le 3.5$ in Fig.~\ref{chi2} (right) indicates a fairly
sharp increase of $\chi^2/(\#\,{\rm d.o.f.})$ around
$\Lambda^{(0)}_{\overline{\rm MS}}\,\rho\approx 0.42$,
corresponding to  our previous default cut-off $\rho^{(0)}_{\rm
max}\approx 0.35$ fm. As a semi-quantitative upper limit for $\rho^{(0)}_{\rm
max}$ one may use the value, where the probability corresponding to
$\chi^2/(\#\,{\rm d.o.f.})$ has dropped from\footnote{Clearly, the
absolute value of the probability is not very meaningful, since, for the
three lattice data sets used, the unknown but significant systematical
errors have been ignored.} $\mathcal{O}(20)\%$ to
$5\%$. In this way, we find 
\begin{equation}
\Lambda^{(0)}_{\overline{\rm MS}}\,\rho^{(0)}_{\rm max}\ = 0.42\div0.435,
\hspace{2ex}\mbox{\ implying}\hspace{3ex} \frac{Q^\prime_{\rm
min}}{\Lambda^{(n_f)}_{\overline{\rm MS}}}= 28.2\div30.8.  
\end{equation} 

As to the $R/\rho$-dependence of the $I\overline{I}$-distance
distribution in Fig.~\ref{lattice} (top, right), a naked-eye estimate
for the uncertainty in $(R/\rho)_{\rm min}$ gives the range
\begin{equation}
\left(\frac{R}{\rho}\right)_{\rm min} = 1.0\div
1.05,\hspace{2ex}\mbox{\ implying}\hspace{3ex} 
x^\prime_{\rm min}=0.31\div0.35.
\label{xprime}
\end{equation}
More quantitative methods do not seem worthwhile here, due to
remaining theoretical ambiguities~\cite{rs-lat}: The integrations over the
$I$- and $\overline{I}$-sizes in the $I\overline{I}$-distance
distribution (c.\,f. Eq.~(\ref{cs})) imply
significant contributions also from larger instantons with $0.35 {\rm \
fm}\lwig\rho,\overline{\rho}\lwig 0.6$ fm, say. 

{\it 5.} Let us now turn to  the crucial question to which extent the
H1 analysis may be directly compared to our 
theoretical predictions. As mentioned earlier,  possibly important fiducial
cuts are lacking in the H1 data. Firstly, we address the fact that 
there is presently no cut on $x^\prime$ in the
data, while these are compared to our QCDINS predictions
with the default $x^\prime$-cut implemented. 

A glance at the lattice data for the $I\overline{I}$-distance distribution
(Fig.~\ref{lattice} (top, right)) reveals that actually,
instanton-effects seem to be very strongly suppressed, as soon as instantons
and anti-instantons  start ``touching'' each other, i.\,e. for small enough
separation, $R/\rho \le \mathcal{O}(1)$. This rapid onset of instanton
suppression corresponds to $x^\prime\lwig 0.35$, which happens to
coincide with our default $x^\prime$-cut. Hence, since
our QCDINS predictions, {\it with} the $x^\prime$-cut
implemented, model quite well the actual suppression
(c.f. Fig.~\ref{lattice} (top, right) and (bottom, right)),   
we consider the lacking experimental
reconstruction of $x^\prime$ not a 
serious problem in principle. 

\begin{figure}
\begin{center}
\parbox{5.1cm}{\includegraphics*[width=5.1cm]{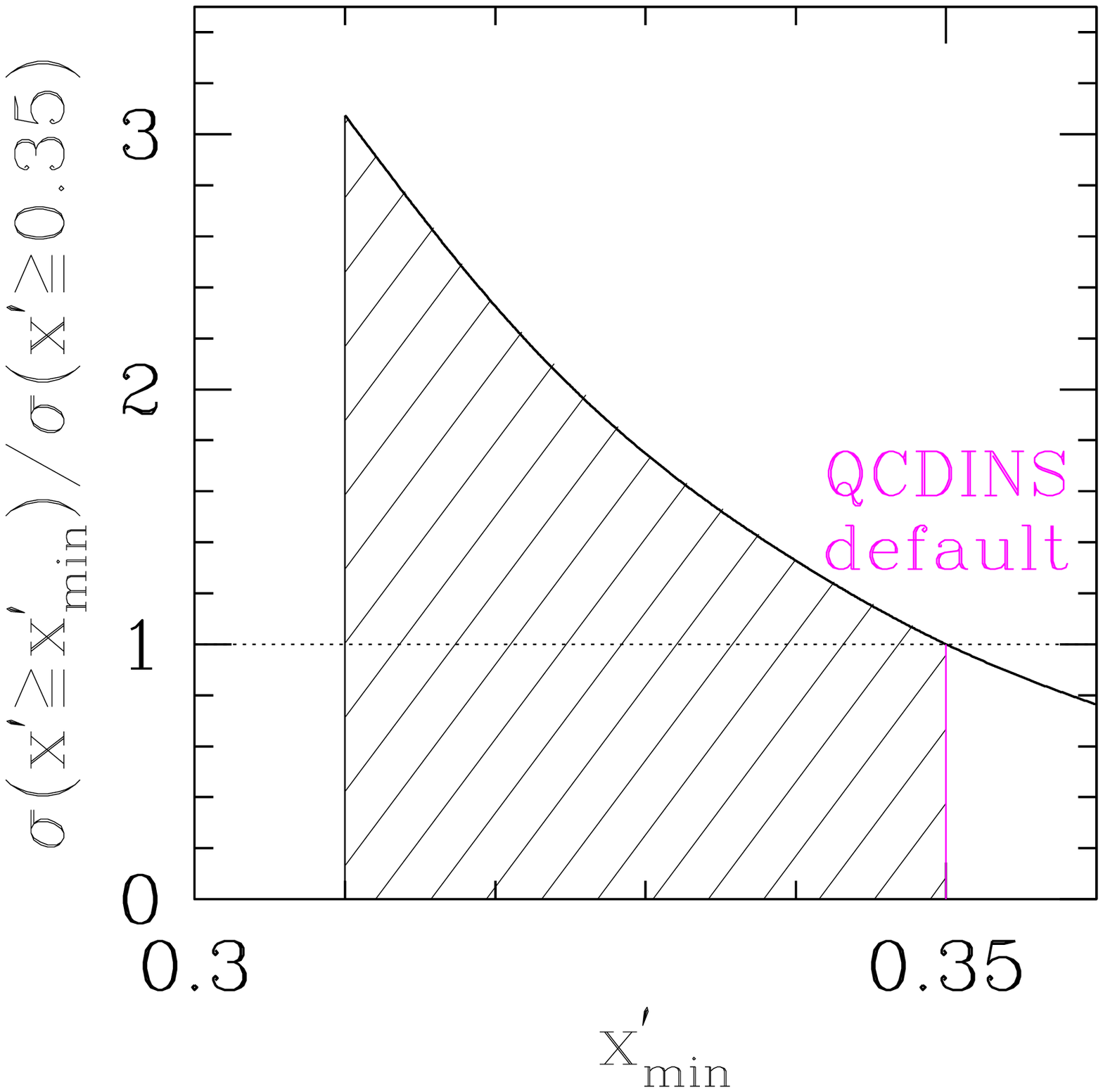}}\hspace{2ex}
\parbox{5.8cm}{\includegraphics*[width=5.8cm]{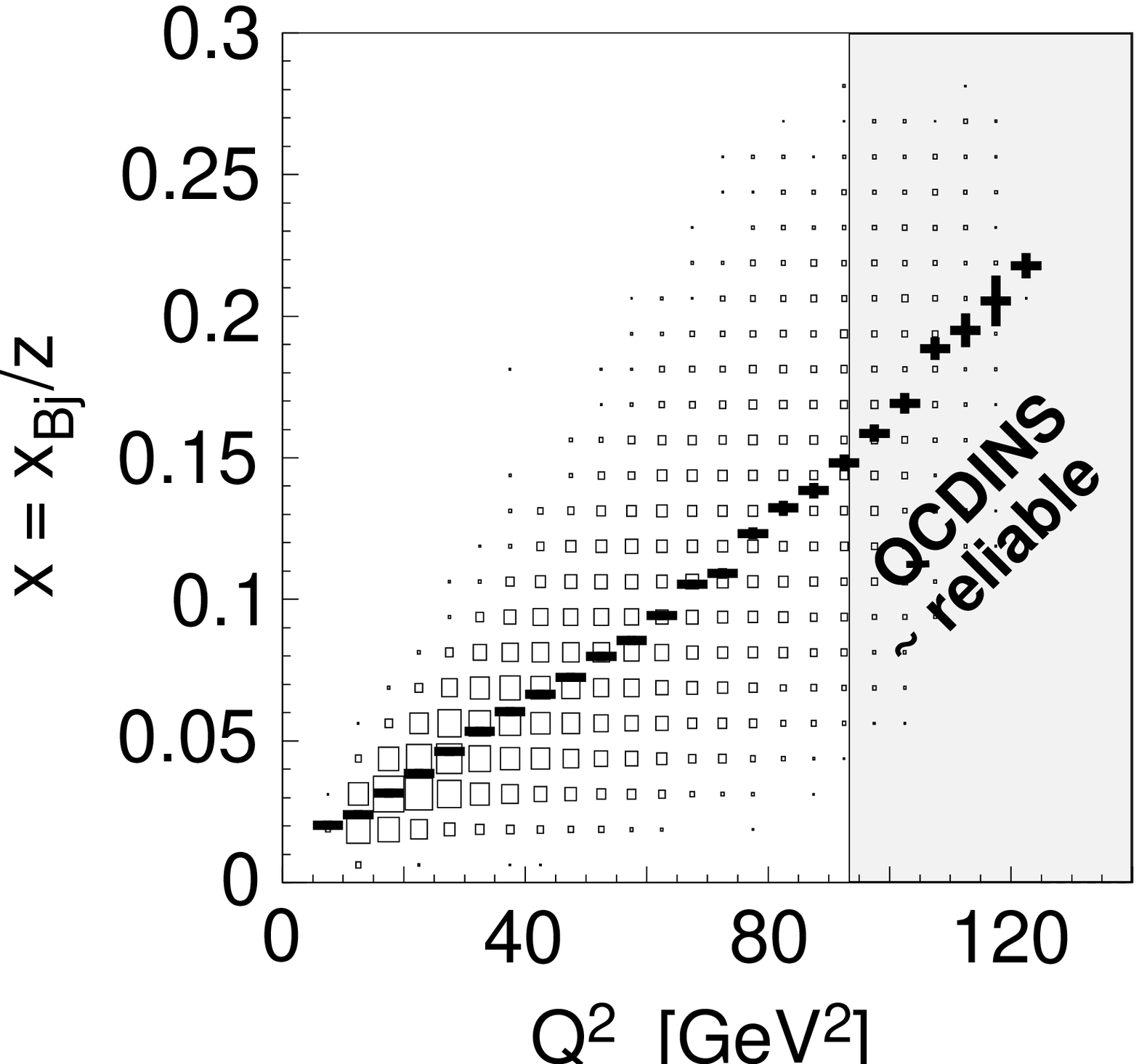}}\hspace{2ex}
\parbox{5.45cm}{\includegraphics*[width=5.45cm]{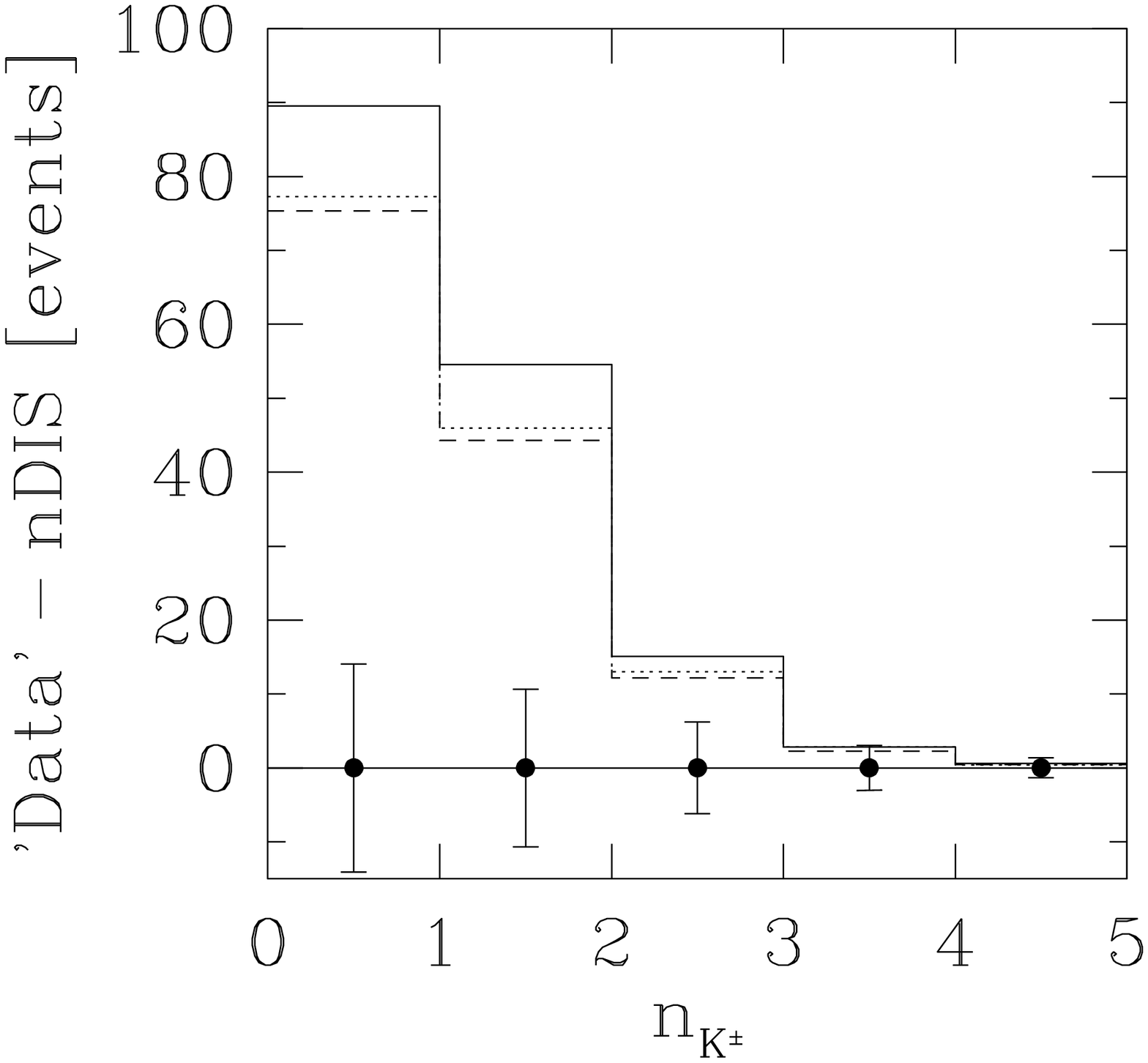}}
\caption[dum]{Left: Common enhancement factor for the instanton signal
in all six H1 observables (Table~\ref{obs}, Fig.~\ref{h1-obs},
cut-scenario {\bf C}), arising 
upon variation of the $x^\prime$-cut within the 
allowed uncertainty window (\ref{xprime}). Middle: Illustration of the
strong $x - Q^2$-correlation according to QCDINS for cut-scenario {\bf
C}. The strong concentration of events around 
$(x\approx 0.035,\,Q^2\approx 20 {\rm \ GeV}^2)$ 
is apparent. Right: Predicted excess
of charged kaons due to 
instanton-induced events according to QCDINS~\cite{qcdins}, 
relative to the normal DIS expectations (nDIS).  
The meaning of the various lines, the statistical errors, cuts and
luminosity are as in Fig.~\ref{h1-obs}. \label{mixed}}   
\end{center}
\end{figure}

Quantitatively speaking, however, the lacking experimental
$x^\prime$-cut leads to a quite substantial ambiguity 
in the expected {\it size} of the instanton signal in the
various observables. In this case, the onset of the actual instanton
suppression from Fig.~\ref{lattice} (top, right), modelled
by an $x^\prime$-cut within the allowed theoretical $x^\prime_{\rm
min}$-uncertainty window (\ref{xprime}), gives rise to a possible
increase of the size of the 
instanton signal up to a factor of three compared to our default
prediction. This is displayed in Fig.~\ref{mixed} (left). Yet, the
predicted {\it shapes} of the six distributions in Fig.~\ref{h1-obs}
are virtually unaffected by this important 
remaining uncertainty. Upon variation of $x^\prime_{\rm min}$ within
the window (\ref{xprime}), we note that to very good approximation,
the predicted instanton-induced excess in all six H1 observables of
Fig.~\ref{h1-obs} varies by the {\it common} factor displayed in
Fig.~\ref{mixed} (left) for cut-scenario {\bf C}.  
In particular, a slight reduction, $x^\prime_{\rm min} =0.325$, compared
to the QCDINS default $x^\prime$-cut, leads to an increase of the
instanton signal by a factor of two, representing a remarkably good
description of the observed excess in four ($n_{\rm ch,\,b}$, ${\rm Sph}$,
$Q^{\prime\,2}_{\rm rec}$ and $1-E_{\rm out\,,b}/E_{\rm in\,,b}$) out
of the six 
experimentally considered observables. While the size of the observed excess in
the remaining two observables  $Et_{\rm b}$ and $Et_{\rm jet}$ is
quite satisfactorily described as well, both $\langle Et_{\rm
b}\rangle$ and $\langle Et_{\rm jet}\rangle$ appear to be shifted
towards smaller values than predicted by QCDINS, and the widths of the
experimental distributions are also considerably narrower.

We shall argue next that this apparent discrepancy is actually a
consequence of the experimentally lacking but theoretically required
cut in $Q^2$, 
\begin{equation}
Q^2\ge Q^2_{\rm min}\approx
Q^{\prime\,2}_{\rm min}=\mathcal{O}(100) \mbox{\rm \ GeV}^2. 
\label{q2cut}
\end{equation}
As a brief reminder~\cite{mrs,qcdins}, this cut 
assures in particular the dominance of ``planar'' handbag-type graphs in
$\sigma^{(I)}_{\rm HERA}$ and all final-state observables. 
The non-planar contributions do not share the simple, probabilistic
interpretation of the planar ones, involve instantons with a  size
determined by $1/Q$ rather than $1/Q^\prime$  and are 
both hard to calculate and hard to implement in a Monte Carlo generator.
On account of their known power suppression in $1/Q^2$ and a
cross-check in the simplest case without final-state
gluons~\cite{mrs}, they can be safely neglected upon application of
the cut (\ref{q2cut}). Because of these reasons, the non-planar
contributions are not implemented in QCDINS. 
  
In order to gain some further insight as to the influence of this
$Q^2$-cut, consider 
$Et_{\rm jet}$, for example, for which one may easily derive the
following simple expression on the parton level in the hadronic CMS,
\begin{equation}
Et_{\rm jet} =
Q^\prime\sqrt{1-\frac{x}{x^\prime}}\sqrt{1-\frac{x}{x^\prime}
\frac{Q^{\prime\,2}}{Q^2}},
\label{etjet}
\end{equation}
involving besides $Q^2,\,Q^{\prime\,2}$ and $x^\prime$, the Bjorken-$x$ of the 
$\gamma^\ast\,g\Rightarrow X$ subprocess in Fig.~\ref{kin-var}.
The important issue to be exploited next is the strong correlation between    
$Q^2$ and $x$, apparent in Fig.~\ref{mixed} (middle).
Already at this point, it is clear that via
its $x$-dependence~(\ref{etjet}), the
$Et_{\rm jet}$-distribution is strongly dependent on the
underlying $Q^2$-spectrum. In absence of the theoretically required
cut in $Q^2$, the $x$-distribution peaks sharply 
around $\langle x\rangle \approx 0.035$ according to QCDINS and
involves a fairly small
$\langle Q^2\rangle\approx 20\div30$ GeV$^2$ (c.\,f. 
Fig.~\ref{mixed} (middle)). 
In fact,
this dominating and theoretically uncontrollable low-$Q^2$ portion of the
spectrum is responsible for the $Et_{\rm jet}$-peak being located at
too large a value, since 
\begin{equation}
\mbox{\rm with\ }\hspace{3ex}\left\{ 
\begin{array}{lclcl}
\langle x^\prime\rangle &\approx& x^\prime_{\rm  min}&=&0.35\\
\langle Q^{\prime\,2}\rangle &\approx& Q^{\prime\,2}_{\rm min}&\approx& 113\
{\rm GeV}^2\\ 
\end{array}
\right\}\hspace{2ex}  
\stackrel{\rm Eq.~(\ref{etjet})}{\Rightarrow}\hspace{2ex}\ \langle Et_{\rm jet}
\rangle  = \mathcal{O}(6) {\rm\ GeV},  
\end{equation}
in good agreement with the direct (but unreliable) result from QCDINS in
Fig.~\ref{h1-obs}, not involving a cut in $Q^2$.
\begin{figure} [tb] 
\vspace{-2ex}
\begin{center}
\parbox{16cm}{\includegraphics*[angle=-90,width=16cm]{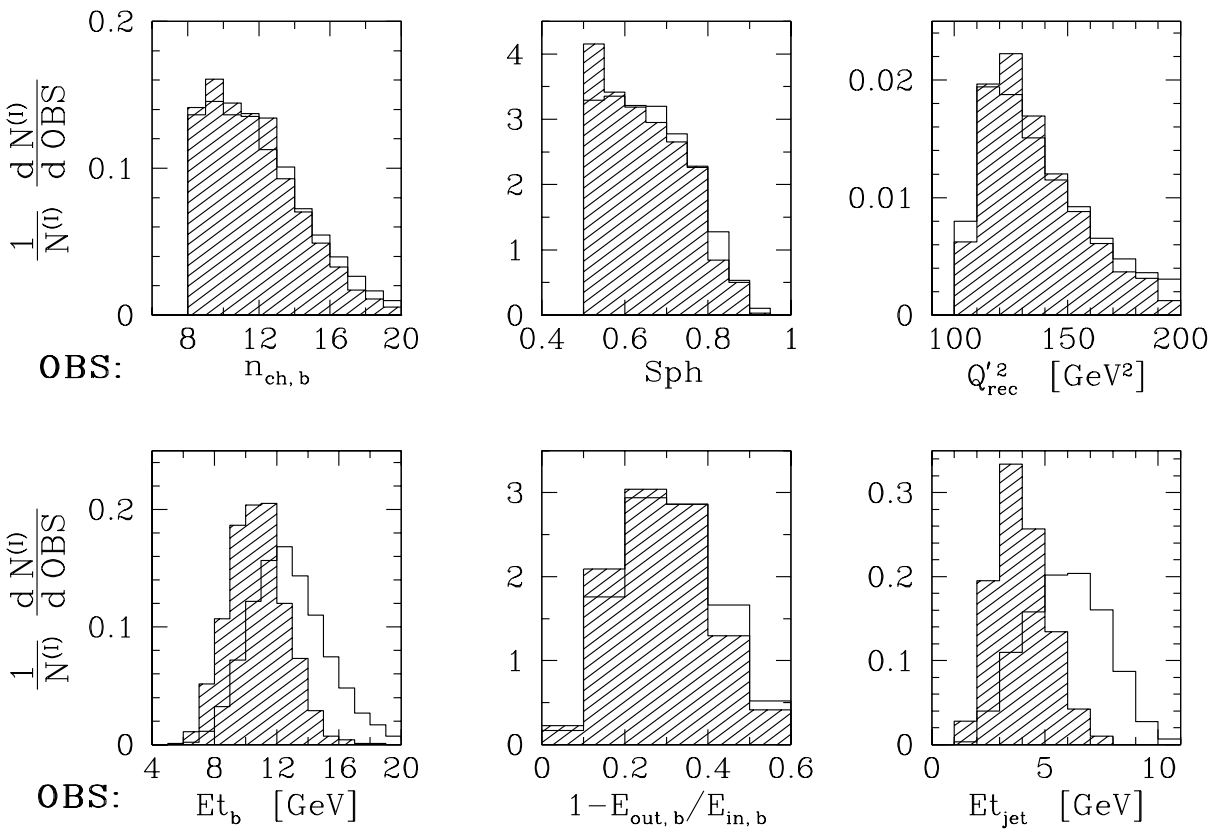}}
\caption[dum]{Dependence of the  
shape-normalized distributions of the six H1 observables 
(Table~\ref{obs}, Fig.~\ref{h1-obs}) for cut-scenario {\bf C} on the
low-$Q^2$ portion of 
the spectrum, for which QCDINS is not reliable. The shaded
distributions are obtained with an additional cut $x\ge 0.15$,
depleting the low-$Q^2$ regime. Note the resulting substantial shift and
narrowing of the $Et_{\rm jet}$ and $Et_{\rm b}$
distributions, with the remaining ones being largely unaffected. 
\label{obsdev} }
\end{center}
\end{figure}

With this insight, let us perform the following exercise. We deplete
the low-$Q^2$ portion of the spectrum by means of a cut, $x\ge 0.15$,
in the (internal) Bjorken-$x$ variable. The QCDINS results should now
be about reliable since from Fig.~\ref{mixed} (middle) we have
effected a shift to $\langle Q^2\rangle\approx 95$ GeV$^2$. The
striking result of this exercise is shown in Fig.~\ref{obsdev} for the
six shape-normalized distributions of the H1 observables. First of
all, we observe that the 
shapes of the four distributions of $n_{\rm ch,\,b}$, ${\rm Sph}$,
$Q^{\prime\,2}_{\rm rec}$ and $1-E_{\rm out\,,b}/E_{\rm in\,,b}$, for
which there was good agreement with the observed excess before
applying the $x$-cut, remain essentially unaffected. Their shapes are thus
rather insensitive to the $Q^2$-spectrum and should be reliably
predicted by QCDINS, even without application of the required cut in
$Q^2$. On the contrary, the shape-normalized distributions of $Et_{\rm
jet}$ and $Et_{\rm b}$ appear both shifted significantly towards
smaller values and are now much narrower. Notably, the
shape-normalized $Et_{\rm jet}$-distribution is now in virtually
perfect agreement with the corresponding experimental quantity. Also
the $Et_{\rm b}$-distribution matches well the experimental trend,
after this depletion of the low-$Q^2$ region.  

On the one hand, these arguments may well serve as indication
that upon proper implementation of the required fiducial cuts the good
description of the data will persist. On the other hand, they should be taken
as a strong encouragement to try and implement the lacking
cuts in $Q^2$ and $x^\prime$ into the data. The essential, remaining
problematics is that searches for an instanton-induced excess in DIS have
to rely on background estimates from nDIS Monte
Carlo generators, in a region of phase space where their accuracy is
not too well known. 

We close with a reminder of our predictions for the rate of
charged kaons that can and should be 
investigated by means of $dE/dx$ in the instanton-enhanced data
sample (c.\,f. Fig.~\ref{mixed} (right) for cut-scenario {\bf C}).      
An excess of kaons, as direct consequence of the flavour democracy of
the instanton-induced interaction, represents a crucial independent
signature. Respective data will hopefully be included in the final H1 analysis.

\section*{Acknowledgements}
We wish to thank our experimental colleagues Volker Blobel for advice
on statistical methods, Tancredi Carli for many helpful discussions and
active collaboration and Eckhard Elsen for a careful reading of the
manuscript.

\end{document}